\documentclass[
aps
,twocolumn
,floatfix
,prl
,superscriptaddress
,longbibliography
]{revtex4-1}

\usepackage[dvips]{graphicx}
\usepackage{bbm}
\usepackage{amsmath}
\usepackage{amsfonts}
\usepackage{lipsum}
\usepackage{bm}
\usepackage[
	colorlinks=true,
	allcolors=blue
]{hyperref}
\usepackage{ulem}
\usepackage{braket}
\usepackage{xcolor}

\begin{document}
	
\pagenumbering{arabic}

\title{Fractional transconductance via non-adiabatic topological Cooper pair pumping}

\author{Hannes~Weisbrich}
\affiliation{Fachbereich Physik, Universit{\"a}t Konstanz, D-78457 Konstanz, Germany}
\author{Raffael~L.~Klees}
\affiliation{Institute for Theoretical Physics and Astrophysics and W{\"u}rzburg-Dresden Cluster of Excellence ct.qmat, Julius-Maximilians-Universit{\"a}t W{\"u}rzburg, D-97074 W{\"u}rzburg, Germany}
\author{Oded~Zilberberg}
\affiliation{Fachbereich Physik, Universit{\"a}t Konstanz, D-78457 Konstanz, Germany}
\author{Wolfgang~Belzig}
\email{Corresponding author: wolfgang.belzig@uni-konstanz.de}
\affiliation{Fachbereich Physik, Universit{\"a}t Konstanz, D-78457 Konstanz, Germany}

\begin{abstract}
	Many robust physical phenomena in quantum physics are based on topological invariants arising due to
	intriguing geometrical properties of quantum states. 
	Prime examples are the integer and fractional quantum Hall effects that demonstrate quantized Hall conductances, associated with topology both in the single particle and the strongly correlated many-body limit.
	Interestingly, the topology of the integer effect can be realized in superconducting multiterminal systems, but a proposal for the more complex fractional counterpart is lacking. 
	In this work, we theoretically demonstrate how to achieve fractional quantized transconductance in an engineered chain of Josephson junctions.
	Crucially, similar to the stabilization of the conductance plateaus in Hall systems by disorder, we obtain
	stable transconductance plateaus as a result of non-adiabatic Landau-Zener transitions.
	We furthermore show that the fractional plateaus are robust to disorder and study the optimal operation regime to observe these effects.
	Our proposal paves the way for quantum simulation of exotic many-body out-of-equilibrium states in Josephson junction systems.
\end{abstract}

\date{\today}
\maketitle

Topology in modern condensed matter physics plays a key role in explaining robust quantized transport properties. 
For instance, the 2D quantum Hall effect is the celebrated archetype of topological physics in condensed matter. 
Here, a 2D electron gas under the effect of a strong magnetic field yields a quantized Hall conductance that is a multiple integer of $e^2/h$ \cite{klitzing1980new,TKKN}, where $e$ is the elementary charge and $h$ is the Planck constant. 
The quantization is extremely precise even under inaccessible microscopic details, such as disorder, impurities, or other unknown details of the experimental configuration, with crucial implications to metrology~\cite{weis2011metrology,gobel2019new}. 
This robustness stems from the nonlocality of topology, where the quantum Hall conductance is proportional to an integer-valued topological invariant, namely the Chern number \cite{hatsugai1993chern}.
%

The idea of quantized charge transport in topological systems was also transferred successfully from 2D systems, such as quantum Hall systems, to 1D topological charge pumps \cite{thouless1983quantization,niu1984quantised,fu2006time,meidan2010optimal,meidan2011topological,ozawa2014anomalous}, where a superlattice potential is varied adiabatically to transport charge along the 1D system. 
Similar to the 2D case, an integer-quantized charge is transported for each pump cycle, with a quantization stemming from the Chern number of related 2D quantum Hall systems \cite{kraus2012topological}. 
Consequently, charge transport in such systems has similar topological robustness to fluctuations and microscopic deviations. 
The idea of topological charge pumps was also realized in recent experiments using ultracold atoms \cite{lohse2016thouless,nakajima2016topological,nakajima2021competition} and photonic systems \cite{kraus2012topological,verbin2015topological}. 
In these systems, the pump parameter can be understood as an extra synthetic dimension that reconstructs the 2D topological invariant from a family of corresponding 1D models \cite{ozawa2016,petrides2018,petrides2022}. 

Following a similar idea, topological Josephson matter \cite{riwar2016multi,eriksson2017topological,xie2017topological,xie2018weyl,deb2018josephson,xie2019topological,gavensky2019topological,houzet2019majorana,klees2020microwave,klees2021ground,weisbrich2021second,Weisbrichtensor,xie2022non,Septembre2022,Gavensky2022,riwar2019fractional,javed2022fractional} or topological superconducting circuits \cite{fatemi2021weyl,peyruchat2021transconductance,herrig2022cooper} were proposed, in which synthetic extra dimensions are implemented as superconducting phase differences that control the topological phase of Andreev bound-state energy bands. 
Those states can become topologically nontrivial in terms of Chern numbers that are defined in these synthetic dimensions.
Furthermore, it was proposed that a quantized transconductance in these systems could be observed when incommensurate voltages are applied between different superconducting terminals \cite{riwar2016multi}.
\begin{figure*}
	\includegraphics[width=1\textwidth]{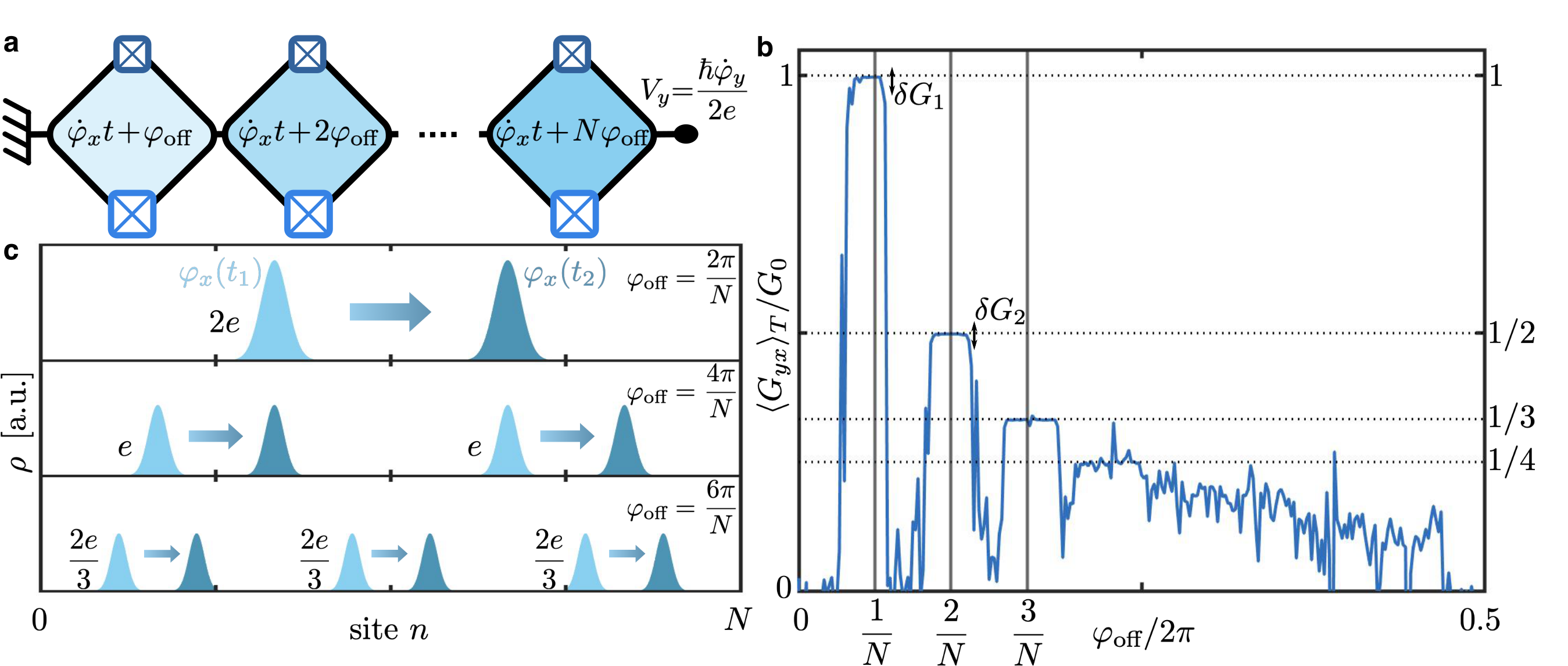}
	\caption{ 
		\textbf{a}, Array of Josephson junctions (blue boxes) arranged as a chain of superconducting loops with fluxes $\varphi_n = \dot{\varphi}_x t + n \varphi_{\text{off}}$ and applied voltage bias $V_y = \hbar\dot{\varphi}_y / (2e)$. 
		\textbf{b}, Transconductance $\langle G_{yx}\rangle_{T} =  \langle I\rangle_{T} /V_x$ in units of $G_{0} = 4e^2 / h$ as a function of $\varphi_{\text{off}}$ averaged over the time $T = 200 \, h / (2e V_y)$ for a chain of length $N=18$ with $2E_{J_1} = E_{J_2}$, $2eV_y=0.01 \, E_{J_1}$, and $\dot{\varphi}_x = -2 \pi \dot{\varphi}_y$, cf.~Eq.~\eqref{eq:trans}.
		\textbf{c}, Local probability density of the excess Cooper pair $\rho(n) = |\braket{n|\psi_1^0}|^2$ of the ground state for different offset fluxes $\varphi_{\text{off}}$ at different times $t_1 < t_2$ given by $\varphi_x(t_1)$ and $\varphi_x(t_2)$, respectively.}
	\label{Fig1}
\end{figure*}
Similar to the 2D quantum Hall effect or 1D topological charge pumps, this quantization occurs in units of $4e^2/h$ and can be linked to the integer-valued Chern number that controls the average transport of Cooper pairs in the system.  
The idea of a Cooper pair charge pump in which single Cooper pairs are transported for each pump cycle was also developed in the framework of Josephson junction arrays \cite{pekola1999adiabatic,aunola2003connecting,leone2008topological,erdman2019fast}.
Furthermore, dual Shapiro steps were recently observed in Josephson junction devices \cite{shaikhaidarov2022quantized,crescini2022evidence} which can be attributed to a quantized transport of Cooper pairs with major significance for quantum metrology.

In 2D electron systems, the interplay between many-body interactions and magnetic fields leads to more exotic topological phenomenon, namely the fractional quantum Hall effect. Here, at fractional fillings of the Landau levels, quantized Hall conductance plateaus appear with fractional values of the conductance quantum $e^2 / h$ \cite{tsui1982two,eisenstein1990fractional,stormer1999fractional,stormer1999nobel}. 
Theoretically, this is explained by the condensation of electrons into new quasiparticles \cite{laughlin1983anomalous,chakraborty1995quantum,jain1990theory,Stern2008,campagnano2012}, so called anyons, that carry a fraction of the electronic charge $e$ and, consequently, have altered anyonic particle statistics that differ from the usual fermionic or bosonic ones. 
The prime example for such anyons are Majorana zero modes that theoretically appear in topological superconductors and promise topologically protected quantum computing based on its anyonic braiding statistics \cite{zilberberg2008,sarma2015majorana,deng2016majorana,aasen2016milestones,karzig2017scalable}.
In conjunction with the single-particle case above,  fractional topological pumps were explored using 1D superlattices with periodically modulated potentials that can be used as synthetic dimensions~\cite{tao1983fractional,bergholtz2008quantum,xu2013fractional,guo2012fractional,budich2013fractional,li2015complete,hu2016fractional}. 
Similar to the fractional ground state of the quantum Hall effect, the ground state in these systems is $\nu$-fold degenerate for fractional fillings of $1/\nu$ and features fractional charge transport \cite{zeng2016fractional,taddia2017topological} along the 1D system. So far, there is nothing comparable to the fractional quantum Hall conductance in superconducting systems.

%
In this work, we theorize a superconducting system that demonstrates fractional topological transconductance. 
Our system facilitates the transfer of a fraction of a Cooper pair along a 1D chain per pumping cycle. 
Unlike adiabatic topological pumps and in similitude to the stabilization of Hall conductance plateaus by disorder, the  fractional transport in our system is stabilized by non-adiabatic Landau-Zener transitions \cite{fedorova2020observation}. This allows for the appearance of fractional plateaus in the transconductance that are robust to various perturbations.
The manuscript is organized as follows: 
First, we discuss the model and the ground-state properties of the chain. We work in the charge-dominated regime, in which the lowest states have at most a single delocalized excess Cooper pair. These states are well separated from excited states with more than one Cooper pair.
Then, we discuss and derive the fractional transport properties when an offset flux along the chain is tuned to periodic points in which, akin to Bohr's quantization condition, a higher periodicity of the ground state can be achieved. 
In the case of finite voltages, however, the fractional transport persists in a sizable region around the high-periodicity operating points, allowing for finite fractional plateaus in the transconductance.
Finally, we analyze the robustness of these plateaus to disorder along the chain.

\section*{Chain of Josephson junctions}
In the following, we consider a chain of Josephson junctions arranged as $N \in \mathbb{N}$ superconducting loops, see Fig.~\ref{Fig1}a. 
Each loop contains two Josephson junctions with Josephson energies $E_{J_1}$ and $E_{J_2}$, and is threaded by a piece-wise increasing flux with $\varphi_n = \dot{\varphi}_x t + n \varphi_{\text{off}}$ threading the $n$-th loop.
Additionally, a voltage $V_y = \hbar \dot{\varphi}_y / (2e)$ is applied to the chain inducing an additional flux at the boundary of the circuit. 
In general, such superconducting circuits are described by the Hamiltonian $H = E_{C} (\hat{\bm{n}}-\bm{n}_g) \bm{c}^{-1} (\hat{\bm{n}}-\bm{n}_g) + U(\hat{\bm{\phi}},\varphi_n)$ \cite{devoret1995quantum,Vool2017}, 
where $E_C = (2e)^2/C$ is the charging energy, which we assume to be homogeneous for simplicity, i.e., $E_{C_j} \equiv E_C$ for all junctions $j$.  
Furthermore, $\hat{\bm{n}} = (n_1,...,n_{N-1})$ are the node charges of the circuit, where $\bm{n}_g = \bm{C}_g \bm{V}_g / (2e)$ is the number of gate charges. The latter is controlled by capacitively coupling the nodes to gate voltages $\bm{V}_g$ with gate capacitances $\bm{C}_g$. The inverse dimensionless capacitance matrix $\bm{c}^{-1}$ and $U(\hat{\bm\phi},\varphi_n)$ summarize the Josephson energies of the circuit, see Methods section for further details.

Importantly, in the limit of $E_C \gg E_{J_i}$, the areas between the superconducting loops act as superconducting islands that are controlled by gate charges $\bm{n}_g$. 
Due to the capacitive coupling between neighboring islands
, the charging energy $E_C$ plays the role of the interaction strength between Cooper pairs on those islands, whereas $\bm{c}^{-1}$ determines the length scale of this interaction. The latter decreases linearly for the circuit shown in Fig.~\ref{Fig1}a.
Consequently, for $\bm{n}_g = \frac{1}{N}$, states with zero and a single excess Cooper pair on the islands are charge-degenerate and allow for a finite supercurrent through the chain. 
These low-energy states are described by the Hamiltonian
\begin{multline}
H_0
=
\sum_{n=1}^{N-1} e^{-i\frac{\dot{\varphi}_yt}{N}}\left(\frac{E_{J_2}}{2}+\frac{E_{J_1}}{2}e^{-i(\dot{\varphi}_xt+n\varphi_{\text{off}})}\right) \ket{n-1}\bra{n} 
\\
+ e^{-i\frac{\dot{\varphi}_yt}{N}}\left(\frac{E_{J_2}}{2}+\frac{E_{J_1}}{2}e^{-i(\dot{\varphi}_x t+N\varphi_{\text{off}})}\right) \ket{N-1}\bra{0} +\text{h.c.} ,
\label{eq:model}
\end{multline}
where $\ket{n}$ is the state with a single Cooper pair on the $n$-th island. 
In this limit, the excess Cooper pair can move between different islands with a hopping amplitude that depends on the applied fluxes and the Josephson energies $E_{J_i}$. 
Excited states of higher energy with two or more Cooper pairs on the islands are gapped out by an energy gap proportional to $E_C$. 

\subsection*{Fractional transconductance}
As we shall see in the following, the voltage biases $2eV_\ell = \hbar \dot{\varphi}_\ell$ ($\ell = x,y$) linearly scan the fluxes over time, leading to topological pumping of the Copper pairs. The quantization of the pumped charge will depend on the offset flux $\varphi_{\text{off}}$.
The main result of this work is the manifestation of a fractional quantization of the transconductance across the device
\begin{align}
	\braket{G_{yx}}_{T \rightarrow \infty} = \frac{C_{xy}}{\nu} \, G_0 
	\label{eq:trans}
\end{align} 
in the long-time average, as shown in Fig.~\ref{Fig1}b. 
Here, $G_0 = 4e^2 / h$ is the conductance quantum for superconducting systems and $C_{xy}=1$ is the first Chern number of the total ground-state band.

\begin{figure*} 
	\includegraphics[width=0.75\textwidth]{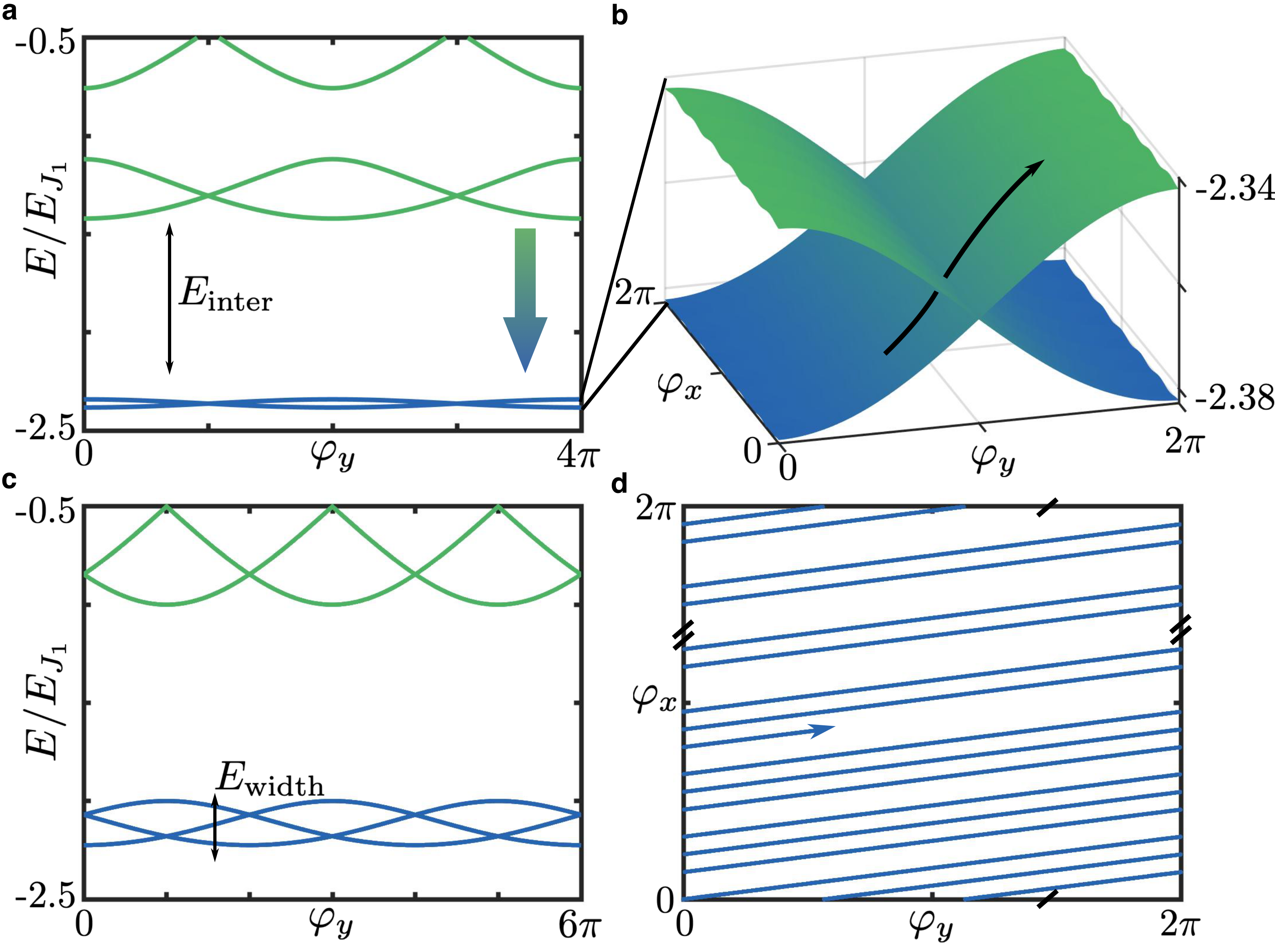}
	\caption{
		\textbf{a}, Ground-state band and first excited states [cf.~Eq.~\eqref{eq:model}] for $\varphi_{\text{off}} = 2\pi \nu / N$ with $\nu = 2$, $N=12$, $E_{J_2} = 2E_{J_1}$, and $\varphi_x=0$. 
		There is an energy gap  $E_{\text{inter}}$ between the ground-state band and other excited states. 
		By changing $\varphi_{\text{off}}$ from $\nu = 2$ to $\nu =3$, one of the excited states is pushed toward the ground-state band (see arrow and panel \textbf{c}). 
		\textbf{b}, Zoom in on the ground-state band for $\nu = 2$ as a function of $\varphi_x$ and $\varphi_y$ with the same parameters as in panel \textbf{a}.
		The adiabatic state evolution is indicated by the black arrow for finite voltages $V_\ell = \hbar \dot{\varphi}_\ell / (2e)$. 
		\textbf{c}, Same as panel \textbf{b} for $\nu = 3$. 
		An additional energy level moved from the excited states toward the ground-state band leading to a $6\pi$-periodicity of the whole band. The width of the lowest band $E_{\text{width}}$ increases with increasing $\nu$.
		%
		%
		\textbf{d}, Applying incommensurate voltages $V_x$ and $V_y$ leads to a non-periodic time evolution $(\varphi_x(t) , \varphi_y(t) )$ for $t \in [0,T]$ that covers the whole torus for long times $T \to \infty$.
		}
	\label{Fig2}
\end{figure*}

To understand the origin of such fractional quantization, let us first qualitatively look at the ground-state of the chain that has a particular form at certain points of the offset flux $\varphi_{\text{off}} =  2\pi\nu / N$ for $\nu \in \mathbb{N}$, see Fig~\ref{Fig2}. 
At $\nu = 1$, there is a single ground state with a single Cooper pair localized in the chain, see Fig.~\ref{Fig1}c. 
The spatial location of the Cooper pair is controlled by the flux $\varphi_x$, which is linearly changed in time via the voltage bias $V_x$. This leads to an adiabatic transfer of this Cooper pair along the chain as long as $2eV_x = \hbar \dot{\varphi}_x \ll E_{\text{inter}}$ is small enough with respect to the gap to the excited states $E_\text{inter}$. 
Increasing the offset flux to $\nu = 2$, the Cooper pair in the ground-state is delocalized on two distinct locations in the chain, see Figs.~\ref{Fig1}c and~\ref{Fig2}a. 
As a function of the linearly changing $\varphi_x$, the delocalized Cooper pair is again transported across the chain.
However, this time and in contrast to the $\nu = 1$ case, the Cooper pair is only transported half the distance due to its bipartite delocalization. 
Similarly, for $\nu = 3$, the ground-state Cooper pair is delocalized at three distinct locations in the chain and the pumped charge is only a third compared to the case $\nu = 1$.
This adiabatic transport is connected with the topological properties of the ground-state band by applying two small incommensurate voltages $V_x = \hbar\dot{\varphi}_x / (2e) = r V_y$ in the spirit of the integer case proposed in Ref.~\cite{riwar2016multi}, where $r \in \mathbb{R}\setminus\mathbb{Q}$.

To understand this result on a quantitative level, we first have a closer look on the specific energy spectrum for an offset flux $\varphi_{\text{off}} = 2\pi\nu / N$. 
For $\nu=1$, there is a single ground state and the instantaneous supercurrent through the chain to first order in the applied voltages is given by $I = 2e \, \partial_{\varphi_y} E_0 /\hbar + 2e \dot{\varphi}_x F_{xy}^0$.
Here, $F_{xy}^0 = i \braket{\partial_{\varphi_x}\psi^0|\partial_{\varphi_y}\psi^0} - i\braket{\partial_{\varphi_y}\psi^0|\partial_{\varphi_x}\psi^0}$ is the Berry curvature of the ground state $\ket{\psi^0}$. 
By applying two incommensurate voltages $V_x$ and $V_y$, the ground state $\ket{\psi^0(\varphi_x,\varphi_y)}$ adiabatically evolves in the periodic space of fluxes $(\varphi_x,\varphi_y)$ according to the second Josephson relation $2e V_\ell = \hbar\dot{\varphi}_\ell$. 
As a result, the time-averaged current through the chain after long averaging times $T \to \infty$ is given by an average over the fluxes $\braket{I}_{T \rightarrow \infty} = \braket{I}_{\varphi_x,\varphi_y}$.
Since $\braket{\partial_{\varphi_y} E_0}_{\varphi_y} = 0$, the only contribution is given by the Berry curvature $\braket{F_{xy}^0}_{\varphi_x, \varphi_y}$, which leads to the integer-quantized transconductance $\braket{G_{yx}}_{T \rightarrow \infty} = G_0 C_{xy}$  in terms of the ground-state Chern number $C_{xy} = \iint_0^{2\pi} F_{xy}^0 d\varphi_x d\varphi_y/ (2\pi) = 1$, cf.~Ref.~\cite{riwar2016multi}.

For $\nu=2$, there are two states in the lowest band with a $4\pi$-periodicity in the $\varphi_y$-direction, see Fig.~\ref{Fig2}a. 
The Chern number of this band is given by $C_{xy} = \iint_0^{2\pi}  \text{tr}(F_{xy}^0)  d\varphi_x d\varphi_y /(2\pi) =1$, where $[F_{xy}^0]_{mm} = i \braket{\partial_{\varphi_x}\psi_m^0|\partial_{\varphi_y}\psi_m^0}-i \braket{\partial_{\varphi_y}\psi_m^0|\partial_{\varphi_x}\psi_m^0}$ is the Berry curvature of the $m$-th eigenstate $\ket{\psi_m^0}$ of the lowest band. 
Applying again a pair of incommensurate voltages, this state will adiabatically evolve in the periodic space of $(\varphi_x,\varphi_y)$ in the lowest band, see Figs.~\ref{Fig2}b and \ref{Fig2}d. 
Hence, due to the increased $4\pi$-periodicity along the $\varphi_y$-axis, there will be an additional factor $1/2$ in the time-averaged current that leads to the fractional transconductance $\braket{G_{yx}}_{\varphi_x,\varphi_y} = C_{xy} G_0/2$. 
By further increasing the offset flux to $\nu=3$, another state will be pushed toward the lowest energy band such that there are three states with a $6\pi$-periodicity along the $\varphi_y$-axis, compare Figs.~\ref{Fig2}b and \ref{Fig2}c. 
Again, this increased periodicity leads to an additional factor $1/3$ in the time-averaged transconductance. 
Overall, by tuning the offset flux $\varphi_{\text{off}}$ such that $\nu \to \nu + 1$, an additional state will be pushed toward the lowest energy band, such that there are $\nu+1$ states in the lowest band with a $2\pi(\nu+1)$ periodicity in $\varphi_y$-direction. 
Correspondingly, the time-averaged transconductance at the particular points $\varphi_{\text{off}} = 2 \pi \nu / N$ yields the quantized fractional values in the adiabatic limit , cf.~Eq.~\eqref{eq:trans} and see the Methods section for the derivation.
Thus, the transconductance of the ground-state band is directly related to the topology of the band and the $2\pi\nu$-periodicity of the state evolution that is controlled by the offset flux $\varphi_\text{off}$ in the chain, see Fig.~\ref{Fig1}b for finite voltages.

	\begin{figure*}
	\includegraphics[width=0.75\textwidth]{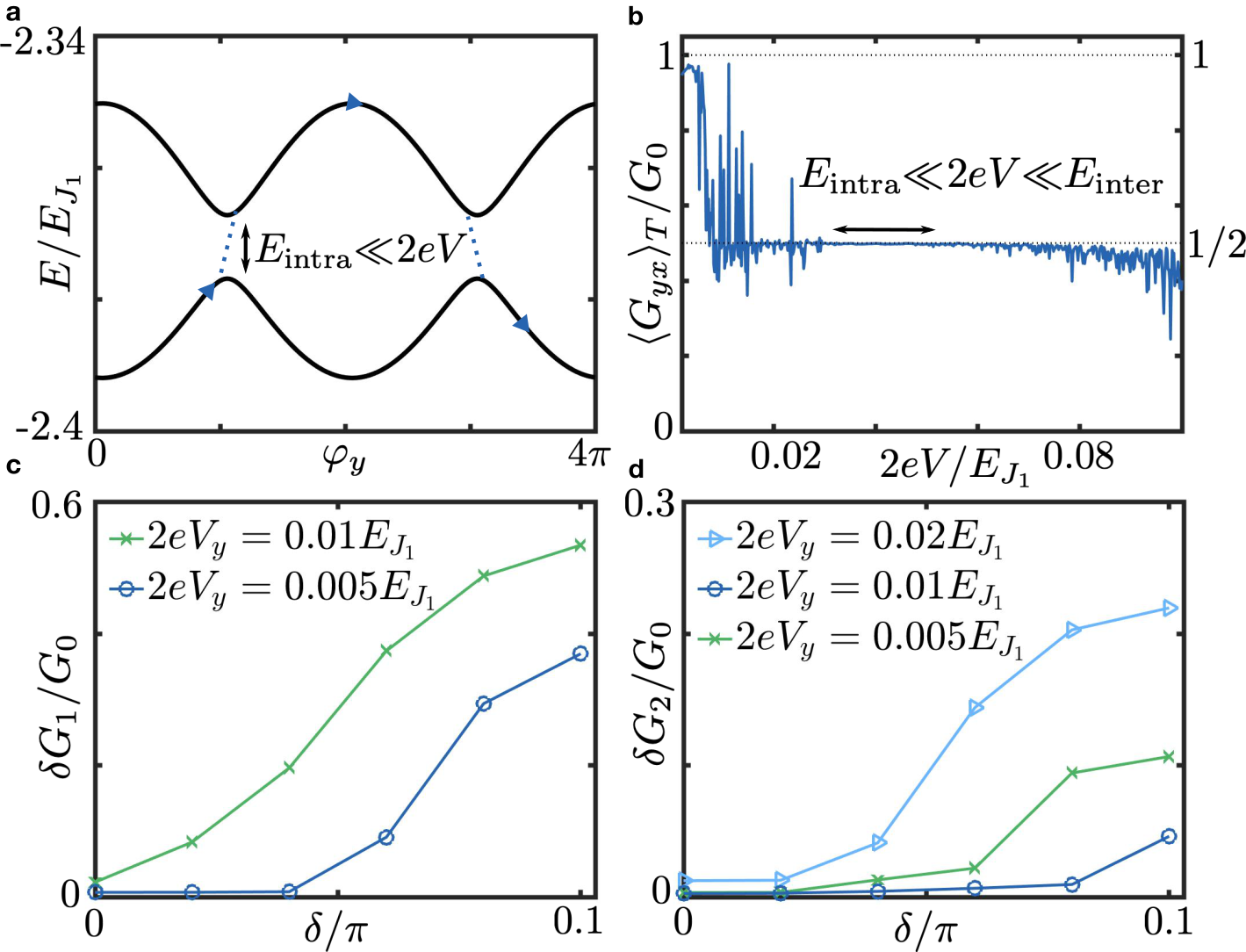}
	\caption{
		\textbf{a}, Lowest band for $\varphi_{\text{off}} = 3.9 \, \pi / N$, $N=12$, $E_{J_2} = 2E_{J_1}$, and $\varphi_x=\pi/2$. 
		At $\varphi_{\text{off}} \neq 2 \pi \nu/ N$, small energy gaps $E_{\text{intra}}$ open between the individual states within the ground-state band. 
		As long as $2eV \gg E_{\text{intra}}$ with $V \equiv \sqrt{V_x^2+V_y^2}$, the state evolution will diabatically jump over these gaps (marked by dots) and maintain the $2\pi\nu$-periodicity (provided that $2eV \ll E_{\text{inter}}$, cf.~Fig.~\ref{Fig2}a).
		\textbf{b},  Transconductance $\langle G_{yx} \rangle_{T} = \langle I \rangle_{T} / V_x$ in units of $G_{0} = 4e^2/h$ as a function of the applied voltage $V = \sqrt{V_x^2 + V_y^2}$ averaged over the time $T = 200 \, h / (2eV_y)$ for $\varphi_{\text{off}} = 3.36\,\pi / N$, $N=12$, $2E_{J_1} = E_{J_2}$, and $V_x = -\pi V_y$. 
		We observe a quantized plateau only in the region where the condition $E_{\text{intra}}(\varphi_{\text{off}}) \ll 2eV \ll E_{\text{inter}}(\varphi_{\text{off}})$ is fulfilled, while the fractional quantization vanishes for too large (small) voltages $2eV \sim E_{\text{inter}}$ $(2eV\sim E_{\text{intra}})$.
		\textbf{c} and \textbf{d}, Standard deviation of the transconductance $\delta G_\nu$ (see Fig.~\ref{Fig1}b) with respect to the quantized values $G_0/\nu$ as a function of the disorder strength $\delta$.
		$\delta G_\nu$ is determined in the range of $\varphi_{\text{off}}$ where quantized plateaus would appear in the case without disorder. 
		The transconductance is averaged over the time $T = 200 \, h/ (2eV_y)$ for $N=18$, $2E_{J_1} = E_{J_2}$, and $V_x=-2\pi V_y$, as in Fig.~\ref{Fig1}b. 
		%
		%
	}
	\label{Fig3}
\end{figure*}

Crucially, we observe quantization plateaus in Fig.~\ref{Fig1}b, which cannot be ascribed to isolated $\varphi_{\text{off}}$ points. Moreover, we consider finite applied voltage biases, which break away from the adiabatic limit that we considered thus far.
Hence, the formation of these plateaus can be only understood beyond the adiabatic approximation. 
In particular, for $\varphi_{\text{off}} \neq 2\pi\nu / N$, there are intraband gaps appearing within the lowest energy band, as shown in Fig.~\ref{Fig3}a for $\varphi_{\text{off}} = 3.9\pi / N$.
The size of these intraband gaps $E_{\text{intra}}$ generally depends on the detuning of $\varphi_{\text{off}}$ away from $\varphi_{\text{off}} = 2\pi\nu / N$, i.e., away from the locations where the intraband gaps vanish. 
These intraband gaps break the increased $2\pi\nu$-periodicity down to a regular $2\pi$-periodicity in the full adiabatic limit. 
%
However, for finite voltages and small intraband gaps, the increased $2\pi\nu$-periodicity can still be recovered for sufficiently diabatic state evolution~\cite{fedorova2020observation}.

In particular, in the limit $E_{\text{intra}} \ll 2eV \ll E_{\text{inter}}$ with $V \equiv \sqrt{V_x^2+V_y^2}$, there will be diabatic Landau-Zener transitions at the avoided crossings within the lowest band, as illustrated in Fig.~\ref{Fig3}a. 
Then, the Landau-Zener probability $P_{LZ} \sim \exp[ -\pi E_{\text{intra}}^2 / (2\hbar\alpha)]$ determines the probability to jump over an avoided crossing \cite{zener1932non}, where $\alpha$ is the slew rate that is proportional to the applied voltage and determines how fast the avoided crossing is traversed.
%
%
Hence, $P_{LZ}\approx 1$ for $E_{\text{intra}}\ll2eV\ll E_{\text{inter}}$ and we restore the $2\pi\nu$-periodicity of the lowest band for detuned $\varphi_{\text{off}}$, which ultimately leads to fractional transconductance plateaus. 
Note, however, that when the applied voltage for a detuned $\varphi_{\text{off}}$ becomes too large, $2eV\sim E_{\text{inter}}$, we will likely observe transitions to excited states within the next energy band that will  destroy the quantization of the transconductance. 
In the other limit, when the applied voltage becomes comparable to the intraband gap, $2eV \sim E_{\text{intra}}$, $P_{LZ} < 1$, adiabaticity yields that there is no $2\pi\nu$-periodicity for the state evolution. 
These two limits can be observed in Fig.~\ref{Fig3}b, where we show the time-averaged transconductance as a function of the voltage bias for a detuned $\varphi_{\text{off}}$.
Thus, clear fractional transconductance plateaus are visible only in regions in which the condition $E_{\text{intra}} \ll 2eV \ll E_{\text{inter}}$ is fulfilled. 
To summarize, the formation of plateaus that are observable in Fig.~\ref{Fig1}b are the result of finite voltages and non-adiabatic Landau-Zener transitions that lead to an increased periodicity of the state evolution for detuned values of the offset flux near $\varphi_{\text{off}} \approx 2\pi\nu / N$.

Note that there are no plateaus for values $\varphi_{\text{off}} \approx 2\pi\nu / N$ if the inter- and intraband gaps are comparable, $E_{\text{intra}} \sim E_{\text{inter}}$, which, for instance, is the case in the example shown in Fig.~\ref{Fig1}b for $\nu > 3$ with $N = 18$. 
In general, the amount of observable plateaus increases with the length $N$ of the chain since the width of the lowest band $E_{\text{width}}$, indicated in Fig.~\ref{Fig2}c, decreases with $N$.
At the same time, however, the plateaus' widths decrease with increasing $N$ because the transitions between different plateaus $\nu \rightarrow \nu+1$ are in a smaller interval of $\varphi_{\text{off}}$ for larger $N$, such that one would observe rather sharp quantized peaks in the transconductance for long chains.

\subsection*{Robustness to disorder}
In addition to the stability of the effect with respect to the offset flux $\varphi_{\text{off}}$, it is also crucial that it is stable under disorder along the superconducting chain. 
Possible sources can be, for instance, non-uniform Josephson energies along the chain or additional disorder in the fluxes of the superconducting loops. 
In the following, we will analyze the robustness of the quantized transconductance with respect to flux disorder $\varphi_n = \dot{\varphi}_x t + n \varphi_{\text{off}} + \delta\varphi_n$, where $\delta\varphi_n$ is the random disorder configuration with zero mean,  $\sum_n\delta\varphi_n=0$, and disorder strength $\delta^2 = \sum_n\delta\varphi_n^2 / N$. 
Note that disorder in the Josephson energies will qualitatively lead to the same outcome, since both sources will lead to disorder in the hopping amplitudes of Eq.~\eqref{eq:model}. 
To quantify the stability of the plateaus in Fig.~\ref{Fig1}b to disorder, we analyze the fluctuations $\delta G_{\nu}$ of the first two plateaus with respect to their fractionally quantized values $G_0 / \nu$. 
If these fluctuations are close to zero, it resembles a perfectly quantized plateau for the whole width in $\varphi_{\text{off}}$ of the disorder-free case. 
Note that even in the clean case, fluctuations are not exactly zero due to finite voltage biases and finite averaging times. 
In the integer case ($\nu=1$ plateau), there is only a single ground state that has a finite gap $E_{\text{inter}}$ to the other excited states. 
This gap is reduced with an increasing disorder strength $\delta$, which leads to a larger probability for transitions to excited states for finite voltages.
However, by applying smaller voltages, we can reduce these transitions and, hence, obtain a better quantization of the integer plateau even with disorder, see Fig.~\ref{Fig3}c. 
Therefore, an increased robustness of the integer plateau to disorder is achieved by simply lowering the applied voltages.

For the fractional plateaus $(\nu \geq 2)$, we observe a fundamental difference to the integer case since there are multiple states in the ground-state band with finite interband gaps $E_{\text{inter}}$.
To achieve the required higher periodicity in the respective state evolution, we need to additionally fulfill the condition $E_{\text{intra}} \ll 2eV$ in order to generate a fractionally quantized plateau in the transconductance. 
Disorder along the chain will also lead to increased intraband gaps $E_{\text{intra}}$ and, in particular, a finite intraband gap even for $\varphi_{\text{off}} = 2\pi\nu / N$.
As shown in Fig.~\ref{Fig3}d, the resulting fractional transconductance is robust against small disorder for intermediate voltages which fulfill $E_{\text{intra}} \ll 2eV \ll E_{\text{inter}}$. 
%
%
For small voltages, the transconductance is sensitive to these increased intraband gaps, even for small disorder, leading to increased fluctuations around the fractional plateaus. 
For larger voltages, we obtain increased transitions to excited states, as the energy gap $E_{\text{inter}}$ is effectively reduced by disorder, which increases fluctuations of the transconductance around the fractional plateaus already for small disorder. 
To conclude, the precise choice of the applied voltage is crucial to achieve optimal robustness against disorder.
For the integer plateau, a lower voltage is beneficial, while an intermediate voltage with respect to the inter- and intraband gap represents the optimum to observe well-pronounced transconductance plateaus at fractions of the conductance quantum.

\section*{Discussion}
In summary, we have shown that quantized transconductance plateaus at fractions of the conductance quantum can be observed in a chain of Josephson junctions. 
This transport property arises due to the interplay between nontrivial topology in the space of fluxes threading the superconducting loops and an increased periodicity of the ground-state evolution.
The fractional transconductance appears in a step-like pattern as a function of the offset flux along the chain and constant plateaus are visible due to non-adiabatic Landau-Zener transitions in the system. 
Crucially, the effect is stable against small disorder due to the diabatic transitions, motivating the experimental exploration of our model in realistic devices. 

The model presented in this work was discussed in detail in the limit when the charging energy is much smaller than the Josephson energy.
For larger charging energies, increased intraband gaps will appear since the single Cooper pair states of the lowest band will hybridize with states that have more than one Cooper pair on the superconducting islands. 
For small intraband gaps, this effect is comparable to the case of weak disorder. 
Furthermore, also dissipation can play a crucial factor in Josephson junction arrays.
Interestingly, interband dissipation toward the lowest band could be actually helpful to suppress transitions to excited states. 
Intraband dissipation in the fractional cases, however, would be harmful for the fractional quantization since it ultimately destroys the increased periodicity of the state evolution of the lowest band. 
Nevertheless, our study pinpoints realistic conditions where fractional transconductance can be observed. Our result paves the way for future experimental realizations of stable fractional transport in superconducting Josephson junction arrays.

\section*{Methods}
\subsection*{Derivation of the low-energy Hamiltonian}
We follow the Lagrangian approach \cite{devoret1995quantum,Vool2017} to derive an effective Hamiltonian for the superconducting chain depicted in Fig.~\ref{Fig1}a. Here we additionally assume that each site is capacitively coupled to a gate voltage  $V_{g,n}$ that can be used to control the gate charges of each site.
The Lagrangian is then given by
\begin{align}
	\mathcal{L} &= \frac{\Phi_0^2}{2}  (C_1+C_2 ) \left(\sum_{n=1}^{N-2} 
	\left(\dot{\hat{\phi}}_{n+1} - \dot{\hat{\phi}}_n \right)^2 
	+ \dot{\hat{\phi}}_1^2 + \dot{\hat{\phi}}_{N-1}^2 \right) \nonumber\\
 &+ \frac{\Phi_0^2}{2} \sum_{n=1}^{N-1}C_{g,n}\left(\dot{\hat{\phi}}_{n}-V_{g,n}\right)^2
	 - U_J,
	\label{eq:lagrangianchain}
\end{align}
with $\bm{\hat{\phi}}=(\hat{\phi}_1,...,\hat{\phi}_{N-1})$ and $\hat{\phi}_{n}$ being the node phases of the circuit. For the sake of simplicity we will assume in the following $C\equiv C_1=C_2$. Additionally, according to Ref.\,\cite{you2019circuit,riwar2022circuit} one has to choose a certain gauge for a time-dependent flux to avoid additional terms in the Hamiltonian, hence, the total Josephson potential $U_{J}$ of the chain takes the form
\begin{align}
	&-U_J = 
	E_{J_2} \cos\left(\hat{\phi}_1+\frac{\varphi_1(t)}{2}-\frac{\varphi_y(t)}{N}-\frac{\sum_j\varphi_j(t)}{2N}\right)\nonumber\\
 &+ E_{J_1}\cos\left(\hat{\phi}_1-\frac{\varphi_1(t)}{2}-\frac{\varphi_y(t)}{N}
 -\frac{\sum_j\varphi_j(t)}{2N}\right) \nonumber\\
		&+ E_{J_2} \sum_{n=2}^{N-1} \cos\left(\hat{\phi}_{n}-\hat{\phi}_{n-1}+\frac{\varphi_{n}(t)}{2}-\frac{\varphi_y(t)}{N}-\frac{\sum_j\varphi_j(t)}{2N}\right) \nonumber\\
	&+ E_{J_1} \sum_{n=2}^{N-1}\cos\left(\hat{\phi}_{n}-\hat{\phi}_{n-1}-\frac{\varphi_{n}(t)}{2}-\frac{\varphi_y(t)}{N}-\frac{\sum_j\varphi_j(t)}{2N}\right)\nonumber\\
	&+ E_{J_2} \cos\left( -\hat{\phi}_{N-1}+\frac{\varphi_N(t)}{2}-\frac{\varphi_y(t)}{N}-\frac{\sum_j\varphi_j(t)}{2N}\right) \nonumber\\
	&+ E_{J_1}\cos\left(-\hat{\phi}_{N-1}-\frac{\varphi_N(t)}{2}-\frac{\varphi_y(t)}{N}-\frac{\sum_j\varphi_j(t)}{2N}\right)\,\label{eq:app_gauge} ,
\end{align}
with $\varphi_y(t) = 2e V_y t / \hbar$ according to the ac Josephson effect and $\varphi_n(t)=\dot{\varphi}_xt+n\varphi_{\text{off}}$ being the flux through the n-th loop. 
Furthermore, $E_{J_1,J_2}$ are the Josephson energies of the two Josephson junctions in a single superconducting loop.
Consequently, the total Hamiltonian can be summarized as
\begin{align}
	\mathcal{H}=\frac{E_C}{2} (\bm{\hat{n}}-\bm{n}_g )^T \bm{c}^{-1} (\bm{\hat{n}}-\bm{n}_g )+U_J ,
	\label{eq:hamiltchain}
\end{align}
where $\bm{c}^{-1}$ is the dimensionless inverse capacitance matrix of the circuit such that $E_{C} = (2e)^2 / (2C)$. 
Here, $\bm{\hat{n}}^T=(\hat{n}_1,...,\hat{n}_{N-1})$ are the node charges and $\bm{n}_g$ the respective offset gate charges that can be controlled by the applied gate voltages $\bm{V}_{g}= 2e \bm{n}_g/\bm{C}_g$ on the respective nodes of the circuit with the gate capacitances $\bm{C}_g$.
Note that the inverse capacitance matrix $\bm{c}^{-1}=C\bm{C}^{-1}$ characterizes the interaction between the Cooper pairs with the respective interaction strength $\sim E_C$. 
In our model, the inverse capacitance matrix consists of the elements
\begin{align} 
	[\bm{c}^{-1}]_{mn}
	=
	\begin{cases}
		\frac{n(N-m)}{N}, & m\geq n\\
		\frac{m(N-n)}{N} , & m<n
	\end{cases}
\end{align}
that results in a long-range interaction between Cooper pairs which linearly decreases with distance in the limit of $C_{g,j}\ll C$.

In the charge-dominated regime $E_{C} \gg E_J$, the charge states of the superconducting islands (i.e., the areas between two superconducting loops) are well defined and the Josephson terms become hopping terms for Cooper pairs between the islands with $2\cos(\phi_i-\phi_j-\gamma(\bm{\varphi}))=e^{-i\gamma(\bm{\varphi})}\ket{n_i,n_j+1}\bra{n_i+1,n_j} + \mathrm{h.c.}$, where $\ket{n_i,n_j}$ is the state with $n_i$ Cooper pairs on island $i$ and $n_j$ Cooper pairs on island $j$, see Ref.~\cite{devoret1995quantum,Vool2017}.
Hence, the low-energy charge states $\ket{0}$ for having zero Cooper pairs on the islands and $\ket{n}$ for having one Cooper pair on island $n$ are described by the Hamiltonian
\begin{align}
	\mathcal{H}=&\sum_{n=0}^{N-1}\varepsilon_n\ket{n}\bra{n}+\sum_{n=1}^{N-1}t_{n}
 \ket{n-1}\bra{n}\nonumber\\&+t_{N}\ket{N-1}\bra{0} + \mathrm{h.c.} ,
\end{align}
where the hopping terms are given by Eq.\,\eqref{eq:app_gauge} and can be simplified to
	\begin{align}
		t_n&= \frac{E_{J_2}}{2}e^{-i(\frac{\dot{\varphi}_y}{N}t+\frac{N+1-2n}{4}\varphi_{\text{off}})}\nonumber\\ 
  &+ \frac{E_{J_1}}{2} e^{-i(\dot{\varphi}_xt+\frac{\dot{\varphi}_y}{N}t+\frac{N+1+2n}{4}\varphi_{\text{off}})}	\end{align}
In addition, the charge energies are 
\begin{align}
	\varepsilon_{j} = \frac{E_C}{2} (\bm{n}_j-\bm{n}_g )^T \bm{c}^{-1}  (\bm{n}_j-\bm{n}_g ) ,
	\label{eq:charging_energies}
\end{align}
with $n_{j} = (0,\ldots ,1,\ldots ,0)$ having a single charge on island $j$ and $n_0 = (0,\ldots,0)$ having no charge on the islands. 
The charge energies can be tuned via $\bm{n}_g$ in such a way that $\varepsilon_j \equiv \varepsilon_{0}$ for all $j$, which represents the charge-degeneracy point of the lowest $N$ states presented here.
The corresponding condition $(\bm{n}_j-\bm{n}_g)^T \bm{c}^{-1}  (\bm{n}_j-\bm{n}_g ) = \bm{n}_g^T\bm{c}^{-1} \bm{n}_g $ for all $j$ can be solved by $\bm{n}_{g} = \frac{1}{N}(1, \ldots , 1)^T$ for small $C_g\ll C$.
Furthermore, the phase of the hopping amplitudes can be simplified by performing a transformation in the time-independent flux $\varphi_{\text{off}}$ with $U=\exp\left(-i(\sum_{n=1}^{N-1}\alpha_n\ket{n}\bra{n})\varphi_{\text{off}}\right)$ with $\alpha_n=\sum_{j=1}^{n}\frac{N+1-2j}{4}$, and $H'=UHU^\dagger$ to find the Hamiltonian given in Eq.\,\eqref{eq:model}.

\subsection*{Fractional transconductance in the adiabatic limit}
To derive the current and the resulting transconductance for small voltage $V_y$ and small $\dot{\varphi}_x$ for $\varphi_{\text{off}}= 2\pi\nu / N$ with $\nu \in \mathbb{N}$, we solve the Schr{\"o}dinger equation $i\hbar\partial_t \ket{\Psi(t)} = H(t) \ket{\Psi(t)}$ by expanding the solution $\ket{\Psi(t)} = \sum_{m,n} c_{m,n}(t) \ket{\psi_m^n(t)}$ in the basis of instantanous eigenstates $\{ \ket{\psi_m^n(t)} \}$ of $H(t)$, with $\ket{\psi_m^n(t)}$ being the $m$-th eigenstate of the $n$-th band with energy $E_m^n(t)$.
Assuming that the initial state is $\ket{\Psi(0)} = \ket{\psi_1^0}$, the differential equation can be solved perturbatively to first order in the phase velocities $\dot{\varphi}_x$ and $\dot{\varphi}_y$ as
\begin{align}
	\ket{\Psi (t) } &= e^{i\theta(t)}\ket{\psi_1^0}\nonumber\\
	&-e^{i\theta(t)}i\hbar\sum_j\dot{\varphi}_j\sum_{(n,m)\neq {(0,1)}}\frac{\braket{\psi_m^n|\frac{\partial\psi_1^0}{\partial \varphi_j}}}{E_m^n-E_1^0}\ket{\psi_m^n}\,,\label{eq:wave}
\end{align}
with the dynamic and geometric phase $e^{i\theta(t)}=\exp(i \int_0^t E_1^0(t') dt' / \hbar )\exp(-\sum_{j = x,y} \int_{\bm{\varphi}(0)}^{\bm{\varphi}(t)} \braket{\psi_1^0|\frac{\partial\psi_1^0}{\partial\varphi_j}} d\varphi_j )$. 
The supercurrent through the right lead can then be computed by $I_y=\braket{\Psi|\hat{I}_y|\Psi}$ with $\hat{I}_{y} = 2e \partial_{\varphi_y} H  / \hbar$.  
Using Eq.~\eqref{eq:wave}, we get
\begin{align}
	I_y(t) &=\frac{2e}{\hbar}\frac{\partial E_1^0(\bm{\varphi}(t) )}{\partial \varphi_y}+2e\dot{\varphi}_xF_{xy}\big(\ket{\psi_1^0(\bm{\varphi}(t) )}\big)\nonumber\\&-2e\sum_{j=x,y}\dot{\varphi}_jK_{yj}\big(\ket{\psi_1^0(\bm{\varphi}(t) )}\big) ,
	\label{eq:current1}
\end{align}
with
\begin{align}
F_{xy}(\ket{\psi_m^n}) = 2\,\text{Im}\Biggl[\sum_{m'}\sum_{n'\neq n}\braket{\frac{\partial\psi_m^n}{\partial \varphi_x}|\psi_{m'}^{n'}}\braket{\psi_{m'}^{n'}|\frac{\partial \psi_m^n}{\partial\varphi_y}}\Biggr]
\end{align} 
being the Berry curvature of the $m$-th eigenstate of the $n$-th band and
\begin{align}
K_{yj}(\ket{\psi_{m,n}})=2\,\text{Im}\Biggl[\sum_{m'\neq m}\braket{\frac{\partial\psi_m^n}{\partial \varphi_y}|\psi_{m'}^n}\braket{\psi_{m'}^n|\frac{\partial \psi_m^n}{\partial\varphi_j}}\Biggr]\,.
\end{align}

For incommensurate phase velocities $\dot{\varphi}_x$ and $\dot{\varphi}_y$, the state is adiabatically swept through the whole space of $(\varphi_x,\varphi_y)$ in the limit of $T \rightarrow \infty$, as proposed in Ref.~\cite{riwar2016multi}.
Hence, the time-averaged current
\begin{align}
	\langle I_y\rangle_{T} = \frac{1}{T} \int_0^T I_y(t') dt'
	\label{eq:finiteAverage}
\end{align}
can be calculated analytically in the long-time limit $T\rightarrow\infty$ as an average over $(\varphi_x,\varphi_y)$ via
\begin{align}
	\langle I_y\rangle_{T\rightarrow \infty}
	&= \langle I_y \rangle_{\varphi_x,\varphi_y}
	= \frac{e \dot{\varphi}_x}{\pi \nu} \Braket{\int_{0}^{2\pi} \text{tr}(F_{xy}^0 ) d\varphi_y }_{\varphi_x} ,
	\label{eq:infiniteAverage}
\end{align}
with $F_{xy}^0$ being the Berry curvature of the lowest band.
Furthermore, we use the periodicity of the states $\ket{\psi_m^0}$ such that 
\begin{align}
	\int_{0}^{2\pi\nu}F_{xy}(\ket{\psi_1^0}) d\varphi_y = 
	\sum_{m \atop E_m^0 \in E^0} \int_{0}^{2\pi} F_{xy}(\ket{\psi_m^0}) d\varphi_y , 
\end{align}
where the sum runs over all states of the lowest band $E^0 = \{ E_m^0  \}$, and 
\begin{align}
	\int_{0}^{2\pi\nu} K_{yl}(\ket{\psi_1^0}) d\varphi_y = 0 .
\end{align}
Finally, averaging the current also over $\varphi_x$ yields the fractional current with the respective Chern number
\begin{align}
	\langle I_y\rangle_{\varphi_x, \varphi_y} &= \frac{e \dot{\varphi}_x }{2\pi^2\nu} \left( \iint_{0}^{2\pi} \text{tr}(F_{xy}^0) d\varphi_xd\varphi_y \right)\nonumber\\
	&= \frac{e \dot{\varphi}_x }{\nu\pi} C_{xy} 
	= G_0 \frac{C_{xy}}{\nu} V_x ,
	\label{eq:avgCurrentTransconductance}
\end{align}
where, in the last step, we used the second Josephson equation $\dot{\varphi}_x\equiv 2eV_x/ \hbar$ that links the applied voltage $V_x$ with the time-dependent flux in the superconducting loop.
Furthermore, $G_0 = 4e^2/h$ is the conductance quantum in superconducting systems.
From Eq.~\eqref{eq:avgCurrentTransconductance}, we can immediately identify the (fractionally) quantized transconductance $G_{yx} = (C_{xy} / \nu) G_0$ with $\nu \in \mathbb{N}$.

\section*{Acknowledgements}
We thank G. Rastelli for helpful discussions. This work was financially supported from the Deutsche Forschungsgemeinschaft (DFG; German Research Foundation) via SFB 1432 (ID 425217212), BE 3803/14-1 (ID 467596333), and Project No. 449653034.

\bibliography{references.bib}
\end{document}